\documentclass[11pt,english]{article}
\usepackage[T1]{fontenc}
\usepackage[latin9]{inputenc}
\usepackage{amsmath}
\usepackage{graphicx,pstricks}
\usepackage{amssymb}
\usepackage[arrow, matrix, curve]{xy}

\usepackage[footnotesize]{caption} 

\makeatletter

\usepackage{epsfig}\usepackage{graphics}

\newtheorem{thm}{Theorem}
\newtheorem{lem}{Lemma}[section]

\newtheorem{defn}[lem]{Definition}

\newcommand{\N}{{\mathbb N}}

\newcommand{\R}{{\mathbb R}}

\newcommand{\E}{{\mathbb{E}}}

\def\N{{\mathbb N}}

\def\R{{\mathbb R}}
\def\P{{\mathbb P}}

\def\E{{\mathbb E}}

\def\0{{\bf 0}}

\def\a{\alpha}

\def\b{\beta}
\def\d{\delta}
\def\e{\varepsilon}

\def\phi{\varphi}
\def\g{\gamma}
\def\l{\lambda}
\def\k{\kappa}

\def\s{\sigma}
\def\t{\tau}

\def\o{\omega}

\def\L{\Lambda}

\def\S{\Sigma}
\def\T{\T}

\def\HH{{\cal H}}
\def\PP{{\cal P}}
\def\NN{{\cal N}}

\def\Cox{\hfill \Box}

\catcode`@=11 \@addtoreset{equation}{section} \catcode`@=12

\usepackage{babel}

\begin{document}


\title{Metastates in mean-field models
 with random\\ external fields generated by Markov chains}
 
\author{M. Formentin
\thanks{Ruhr-Universit\"at Bochum, Fakult\"at f\"ur Mathematik, Universit\"atsstrasse
150, 44780 Bochum, Germany {and
Dipartimento di Fisica ``Galileo Galilei'', Universit\`a degli studi di Padova, Via Marzolo 8, 35131 Padova, Italy}, \texttt{Marco.Formentin@rub.de
} %
},\,
C. K\"ulske 
\thanks{Ruhr-Universit\"at Bochum, Fakult\"at f\"ur Mathematik, Universit\"atsstrasse
150, 44780 Bochum, Germany, \texttt{Christof.Kuelske@rub.de}}, \, and A. Reichenbachs
\thanks{Ruhr-Universit\"at Bochum, Fakult\"at f\"ur Mathematik, Universit\"atsstrasse
150, 44780 Bochum, Germany, \texttt{Anselm.Reichenbachs@rub.de} %
}}

\maketitle

\begin{abstract} 
We extend the construction by K\"ulske and Iacobelli of metastates in finite-state mean-field models 
in independent disorder to situations where the local disorder terms 
are a sample of an external ergodic Markov chain in equilibrium. We show that  
 for non-degenerate Markov chains, 
the structure of the theorems is analogous to the case of i.i.d. variables 
when the limiting weights in the metastate are expressed with the aid of a CLT 
for the occupation time measure of the chain.\\
As a new phenomenon we also show in a Potts example that
for a degenerate non-reversible chain this CLT approximation is not enough, and that 
the metastate can have less symmetry than the symmetry of the interaction 
and a Gaussian approximation of 
disorder fluctuations would suggest.    

\end{abstract}

\bigskip 


\section{Introduction} \label{sect:intro}

Metastates are  useful concepts in analyzing the large-volume dependence  
of disordered systems, both of lattice and mean-field systems. They go back to the constructions 
of Aizenman and Wehr \cite{AiWe90}, and Newman and Stein \cite{NeSt97}.  

In the present paper we continue to analyze a class of mean-field systems, 
moving from a situation where the quenched disorder was assumed to be i.i.d. 
 \cite{CI} to a situation where the disorder is generated by  a Markov chain. 
Our models have a 
finite local state space $E$, and a finite space
$E'$ of possible values of the local disorder variables. 
While in the previous analysis the disorder variables were assumed 
to be i.i.d. over the sites, we now treat a quenched disorder which is correlated and generated by a Markov 
chain with transition matrix $M$. For a study of the Curie-Weiss model in a dynamical external field see \cite{Do, Re}. 

A metastate is a probability measure on the infinite-volume Gibbs measures 
of a disordered system which intuitively gives the likehood to pick a 
Gibbs state in a large but finite volume. For precise definitions see below and refer 
to the books \cite{Bo06, Ne97}. 
Non-trivial metastates can occur when the phenomenon of random 
symmetry breaking appears, meaning that two or more states are equivalent 
on the average, but their symmetry is broken by disorder fluctuations 
in finite volume.  For examples of disordered models with such random symmetry 
breaking see \cite{ BoEnNi99,BoGa98, EnNeSc06,Ku97, Ku98, Ku98b}, for genuine 
applications of metastate idea to spin-glasses see 
\cite{NeSt01, NeSt02} and in particular the groundstate uniqueness result in 2D proved 
in \cite{ArDaNeSt09}. 

Now, in the finite-type mean-field models of  \cite{CI} precise limit theorems for the weights 
in the metastates were derived, combining a multivariate 
CLT  for disorder fluctuations 
and a large-deviation analysis of the spin-part of the model. 
Further a geometric picture was given to identify the {\it invisible states},
namely those  Gibbs states which are always 
overshadowed by disorder fluctuations favoring other competing states.     

In the present paper we ask: What changes if the external randomness comes 
from a Markov Chain? 

Suppose we are dealing with a non-degenerate Markov chain on a finite state-space at first, meaning 
that the covariance matrix $\S_M$ of the limiting Gaussian distribution 
of the occupation time-vector  of the Markov chain $M$ has the full possible rank.   
Then the basic structure of the theorems stays the same, using the same 
geometric construction to identify visible Gibbs states $j$ and their weights $w_j$.  
However, let us stress the main difference to the independent case, which is
that the  $w_j$'s are given in terms of $\S_M$  
and this object 
carries more information about the Markov chain 
then just its invariant distribution $\pi$. Which states 
are Gibbs states depends on $\pi$,  
but the precise form of the weights and  
whether  a given state is visible or not
can vary   over Markov chains having the same $\pi$ 
but different transition matrix $M$. 
This is different from the i.i.d. case where the single-site 
distribution $\pi$ completely specifies the disorder distribution. 

New phenomena on an even more fundamental level 
appear when we allow for Markov chains which are degenerate, i.e., whose
 $\S_M$ does not have full rank. 
We specify to a concrete example and look 
at the 3-state random field Potts model, with $3$-state local random fields 
coupling by delta-interaction in the Hamiltonian to the spins. 
We take $M$ on \mbox{$E'=\{1,2,3\}$} to be doubly stochastic ergodic matrix 
which can be seen as a softened version 
of a permutation matrix,  where one  keeps however one (and only one) 
transition between an ordered pair of states to be deterministic, 
say from state $1$ to $2$. 
Although the equidistribution on the states is invariant and the covariance 
matrix $\S_M$ becomes degenerate but is symmetric 
under exchange of states $1$, and $2$,
the metastate for the random model with disorder generated by the Markov
chain in equilibrium is not symmetric under an 
exchange of the spin labels $1$ and $2$. 
Such a phenomenon is only possible because of two features: 
The almost degeneracy between two Gibbs-states up to finite but non-zero free energy differences in arbitrary large volumes in the possible realizations of the chain, 
and the lack of time-reversal symmetry of the Markov chain. 

The remainder of the paper is organized as follows. 
In Section 2 we review definitions about metastates 
and the constructions from \cite{CI} needed for mean-field models, 
including the geometry of free energy fluctuations 
and derivation of weights for pure states. 
Then we are ready to give the result of the present paper about metastates 
for non-degenerate Markov chains. 
Next we present the results for the 
$3$-state Potts model with quenched disorder generated by a degenerate chain. 
In Section 3 the proofs for non-degenerate Markov chains are given, building 
on results of the previous paper \cite{CI}.  
In Section 4 the proof for the $3$-state Potts model in the degenerate case is given. 
\newpage 

\section{Main Results}

\subsection{Mean-field Spin Models in a random field generated by  a Markov chain}

\subsubsection{The Boltzmann distribution of the spins}
For spin variables $\sigma(i)$, at each site $i\in \{1,\dots,n\}$ 
taking values in a finite set $E$, we consider a mean-field model with Hamiltonian $nF(\nu)$, and Boltzmann distribution given by:
\begin{equation}
\begin{split}\label{1.3.2}
&\mu_{F,n}[\eta(1),\ldots,\eta(n)](\s(1)=\o(1),\dots,\s(n)=\o(n))\cr
&=\frac{1}{Z_{n}[\eta(1),\ldots,\eta(n)]}\exp\left(-nF(L_n^{\o})\right )\prod_{i=1}^n\a[\eta(i)](\o_i)\cr 
\end{split} 
\end{equation}
where $L_n^{\o}=\frac{1}{n}\sum_{i=1}^n \d_{\o(i)}$ is the empirical spin distribution. 
The Boltzmann distribution depends on the quenched disorder $\eta(1),\ldots,\eta(n)$ via the a-priori distribution $\a[b]\in\mathcal{P}(E)$, $b\in E'$, where the type-space $E'$ of the local disorder variables 
is taken to be finite.  Throughout the paper we use the notation $\PP(\hat E)$ for the probability measures 
on a space $\hat E$, not mentioning the sigma-algebra explicitly unless necessary.

\subsubsection{The Markov chain of the disorder variables}
Let the probability distribution of the disorder be given by 
a Markov chain on the finite space $E'$ with transition matrix $M$ and invariant measure $\pi$, 
written in vector notation as $\pi^t M=\pi^t$ where $t$ denotes transposition of 
the column vector $\pi$. 
The chain is supposed to be in equilibrium, i.e. the initial state $\eta(0)$ is chosen according to the invariant distribution of the chain $\pi\in\mathcal{P}(E')$, and 
$\P_{\pi}$ is the corresponding measure on $(E' )^{\N}$, that is the one with  
$\P_{\pi}(\eta(i+1)=b'|\eta(i)=b)=M(b,b')$ and $\P_{\pi}(\eta(i)=b)=\pi(b)$ for all $i\in \N$ 
and $b\in E'$.  We will assume that the transition matrix 
$M$ is ergodic, meaning that there exists 
a power $M^r$ for which all matrix elements are strictly bigger than zero. 

Given the first $n$ realizations of the Markov chain $\eta$ we denote by 
$$\L_n(b)=\{i\in\{1,\dots,n\}: \eta(i)=b\}$$ 
the set of \textit{$b$-like sites}, $b\in E'$. Then the number $|\L_n(b)|$
is the occupation-time of the chain at state $b$ up to ``time'' $n$, and the normalized 
expression 
$$\hat{\pi}_n(b)=\frac{|\L_n(b)|}{n}$$ 
 is the relative frequency of $b$-like sites in the sample. The empirical spin distribution on the $b$-like sites is given by 
 $$\hat{L}_n(b)=\frac{1}{|\L_n(b)|}\sum_{i\in\L_n(b)}\delta_{\omega(i)}$$
and moreover the total empirical spin distribution can be written as the scalar product of $\hat{\pi}_n$ and $\hat{L}_n$, i.e.
$$L_n=\sum_{b\in E'}\hat{\pi}_n(b)\hat{L}_n(b).$$\par\bigskip
The equilibrium states of the system are given by the minimizers of the free energy functional \cite{CI}
\begin{equation}
\hat\nu\mapsto\phi[\pi](\hat\nu)
\end{equation}
where
\begin{equation}
\begin{split}\label{free-energy}
&\phi:\mathcal{P}(E')\times \mathcal{P}(E)^{E'}\rightarrow\R,\cr
&\phi[\pi](\hat\nu)=F\left(\sum_{b\in E'}\hat\pi(b)\hat\nu(b)\right)+\sum_{b\in E'}\hat\pi(b)S\left(\hat\nu(b)|\a[b]\right).\cr 
\end{split} 
\end{equation}
Here, $S(p_1|p_2)=\sum_{a\in E}p_1(a)\log\frac{p_1(a)}{p_2(a)}$ is the relative entropy.\\

\noindent Together with the free energy we consider its linearization at some fixed minimizer $\hat \nu_j$ as a function of $\pi$, which reads as follows

\begin{equation}
\phi[\tilde\pi](\hat\nu_j)-\phi[\pi](\hat\nu_j)=-B_j[\tilde\pi-\pi]+o(||\tilde\pi-\pi ||)
\end{equation}
As in \cite{CI}, throughout the present paper we restrict ourselves to the following non-degeneracy conditions:\label{aaa}
\begin{enumerate}
\item the set $M^*$ of minimizers of the free energy is finite and the Hessian  is positive at each minimizer.
\item No different minimizers $\hat\nu_i$ and $\hat\nu_j$ have the same $B_i=B_j$.
\end{enumerate}
Our first results concern the metastate on the level of the empirical spin-distribution. Denote by $\rho[\eta](n):=\mu_{F,n}[\eta](L_n)\in\PP(\PP(E))$ the distribution of $L_n$ under the finite-volume Gibbs-measure. The concentration of these measures around the finitely many minimizers of the free energy function $\pi M^*=\{\pi\hat{\nu}_j, j=1,\dots,k\}$ holds by assumption. How this concentration takes 
place in asymptotically large volumes 
is made precise by the following definition of a metastate.
\begin{defn} Assume that for every bounded continuous $\psi:\mathcal{P}\left({\mathcal{P}}(E)\right)\times (E')^{\infty}\rightarrow \R$ the limit
\begin{equation}
\lim_{n\rightarrow\infty}\int \P (d\eta)\psi(\rho[\eta](n),\eta)=\int K\left(d\rho,d\eta\right)\psi(\rho,\eta)
\end{equation} 
exists. Then the conditional distribution $\k[\eta](d\rho)=K(d\rho|\eta)$ is called the metastate on the level of the empirical distribution.
\end{defn}
Before stating the theorem we must introduce the definition of the region of stability for the minimizers:
\begin{defn} Let us define the set
\begin{equation}
R_j:=\{x\in T\mathcal{P}(E'),\langle x,B_j\rangle>\max_{k\neq j}\langle x,B_k\rangle\} 
\end{equation}
where $T\mathcal{P}(E')=\{x=(x(1),\ldots,x(|E'|)):\sum_{i=1}^{|E'|}x(i)=0\}$ is the tangent space of field type measures.
We say that $R_j$ is the stability region of $\hat\nu_j$.
\end{defn}

We will use the following CLT for  the occupation time 
measure (see Appendix). 

{\bf Fact:} For an ergodic finite-state Markov chain, the standardized occupation-time fluctuation 
measure 
$\sqrt{n}(\hat\pi_n-\pi)$ converges in distribution, as $n$ tends to infinity,  to a centered 
Gaussian distribution $G$ with a covariance matrix $\S_M$ on the $|E'|-1$ 
dimensional vector space $T\mathcal{P}(E')$. 
\bigskip 

{\bf Warning:} Ergodicity of the Markov chain does not imply that $\S_M$ has the full rank $|E'|-1$, 
we will discuss in detail an example where this is not the case. 
In generalization of the case of i.i.d. $\eta(i)$'s we have the following result. 

\begin{thm}\label{Theorem1} Consider a mean-field model with quenched disorder generated by an ergodic finite state Markov chain $M$  
in equilibrium $\pi$, with full rank occupation-time covariance $\S_M$. 
Suppose the non-degeneracy conditions 1. and 2. on the spin model (see pag.\pageref{aaa}).
Then the metastate on the empirical spin-measure exists and takes the form 
\begin{equation}
 \kappa[\eta](d\rho)=\sum_{j=1}^kw_j\delta_{\delta_{\pi\hat\nu_j}}(d\rho)\mbox{ for $\P_{\pi}$-a.e. $\eta$.}
\end{equation}
The weights are $w_j=\P_{\S_M}(G\in R_j)$ where $G$ is a centered Gaussian 
on $T\mathcal{P}(E')$ with covariance $\S_M$. 
\end{thm}

As in the case for i.i.d. $\eta(i)$'s we also have the analogous version 
for the metastate on the level of spin distributions. 
We call a function on an infinite product of a finite space 
continuous (w.r.t. local topology) if it is a uniform limit of local functions. 
For probability measures on $\PP(E^\infty)$ we use the weak topology (according 
to which a sequence of measures converges iff it converges on continuous test-functions), 
and for $\PP(E^\infty)\times (E')^\infty$, we use the product topology. 

\begin{defn} Assume that, for every bounded continuous $\Xi:\PP(E^\infty)\times (E')^\infty\rightarrow \R$  
the limit 
\begin{equation}
\begin{split}\label{nondeg3}
&\lim_{n\uparrow \infty}\int\P(d\eta)\Xi(\mu_{n}[\eta],\eta)=\int J(d\mu,d\eta)\Xi(\mu,\eta)\cr
\end{split} 
\end{equation}
exists. Then the conditional distribution $\k[\eta](d\mu):=  J(d\mu|\eta)$ 
is called the {\em AW-metastate on the level of the states}. 
\end{defn}

\begin{thm}\label{Theorem2} In the situation of Theorem \ref{Theorem1},
 the metastate on the level of the states equals 
\begin{equation}
\begin{split}\label{AW-metastate}
&\k[\eta](d\mu)=\sum_{j=1}^k w_j \d_{\mu_j[\eta]}(d\mu)
\end{split} 
\end{equation}
with the product measures 
$\mu_j[\eta]:= \prod_{i=1}^\infty \g[\eta(i)](\,\cdot\,| \pi \hat\nu_j)$ 
given by the kernels 
\begin{equation}
\begin{split}\label{colonel}\g[b](a| \nu)=\frac{e^{-d F_{\nu}(a)}\a[b](a)}{
\sum_{\bar a\in E}   e^{-d F_{\nu}(\bar a)}\a[b](\bar a)}
\end{split} 
\end{equation}
where $d F_{\nu}(a)$ is the $a$-th coordinate 
of the differential of the function 
$F$ on $\PP(E)$ taken at the point $\nu \in \PP(E)$.
\end{thm}

The proofs of these theorems are analogous to the i.i.d. case, however 
with some additional features when decoupling 
properties of random field realizations in separate finite volumes need to be treated. 


\bigskip
\bigskip

\subsection{Potts model in a random-field generated by a degenerate Markov chain}

 The random-field Potts model of the form we consider is given by the energy 
$$F(\nu)=- \frac{\b}{2} (\nu(1)^2+\dots+ \nu(q)^2)$$  
where $E\equiv E'=\{1,\dots,q\}$ and the random fields couple locally to the spins 
via the local measures $\a[b](a)=\frac{e^{B 1_{b=a}}}{e^B + q-1}$. 
We will consider a Markov chain $M$ with invariant measure 
$\pi$ given by the equidistribution.  

The structure of the phases depends on $M$ only through $\pi$ 
and the case of i.i.d. equidistributed variables has already been treated 
in \cite{CI}, so let us recall the relevant information from there:
 The total empirical distribution of the minimizers of the free energy functional 
has the form $\nu_{j,u}\in \PP(E)$, with possible order in 
direction $j$,  such that 
$\nu_{j,u}(j)=\frac{1+ u (q-1)}{q}$,  $\nu_{j,u}(i)=\frac{1- u}{q}$ 
for $i\neq j$, where the symmetry-breaking parameter $u$ 
obeys the mean-field equation 
\begin{equation}
\begin{split}\label{rr}
 u &=\frac{e^{\b u}}{ 
e^{\b u} + e^B + (q-2)}
- \frac{1}{
e^{\b u + B} +  (q-1)}
\end{split} 
\end{equation}

In a region of small $B$ and large $\b$ there is a non-zero solution of this 
mean-field equation; to decide whether this or the zero solution corresponds to 
the global minimizer one looks at the corresponding free energy 
which reads 
\begin{equation}
\begin{split}\label{1.3.2}
&\log\frac{e^B+q-1}{e^{\b u}+e^B+q-2}+\frac{\b(q-1)}{2 q}u^2+\frac{\b}{q}u -\frac{1}{q}\log\frac{e^{\b u+B}+q-1}{e^{\b u}+e^B+q-2}
\end{split} 
\end{equation}
There is a curve in the $(\b,B)-$plane 
when there is an equal-depth minimum at $u=0$ and a positive value of $u=u^*(\b,q)$. This 
means that on this curve there is a coexistence of $q$ ordered states 
and one disordered state where $u=0$ in the random model in the very same way 
as  it is happening for the nonrandom model where $B=0$. 

The stability vector for the state $\nu_{1,u}$ is 
\begin{equation*}\begin{split}
\hat B_{\nu_{1,u}}
= \begin{pmatrix}
  \frac{q-1}{q}\log\frac{e^{\b u + B} + q-1}{e^{\b u}  + e^{B} + q-2}\\
   -\frac{1}{q}\log\frac{e^{\b u + B} + q-1}{e^{\b u}  + e^{B} + q-2}\\ 
  \dots \\ 
  -\frac{1}{q}\log\frac{e^{\b u + B} + q-1}{e^{\b u}  + e^{B} + q-2}
\end{pmatrix}\cr
 \end{split}
\end{equation*}
pointing 
 into the direction $1$ in $T \PP(E')$ which is intuitively clear since a majority of 
$1$-type random fields favor the state having a majority of $1$-type spins. 
The other stability vectors are given by permutation. 

Let us specialize to $q=3$.
 In the case of i.i.d. random fields the metastate on the spin-distributions becomes 
$$\k[\eta](d\mu)=\frac{1}{3}\sum_{j=1}^3 \d_{\mu_j[\eta]}$$ 
with  $\mu_j[\eta]= \prod_{i=1}^\infty \g[\eta(i)](\,\cdot\,| \nu_{j,u=u^*(\b,q)})$. 
In particular the state for $u=0$ is invisible.  This follows since the stability 
vector for this state must vanish, by symmetry, as was explained in the previous paper. 

We note that the form of the minimizers and the stability vectors depend only 
on $\pi$. So, equally for a Markov chain $M$ which has the equidistribution as its invariant measure, 
and has a non-degenerate covariance matrix $\S_M$ the state for $u=0$ will 
stay invisible, and the metastate will become 
\begin{equation}\label{Pottsmeta}
\k[\eta](d\mu)=\sum_{j=1}^3 w_j \d_{\mu_j[\eta]}
\end{equation}
 with weights  
$w_1,w_2,w_3$ which are given by the probabilities of finding a (in general non-isotropic) Gaussian
in the corresponding equal-sized stability regions, see (\ref{Pottsweights}).
 Hence in general the $w_1,w_2,w_3$ will be different. 
\bigskip 

The form \eqref{Pottsmeta} changes completely, in that the metastate is now supported on mixed states, when we 
consider, for $p\in (0,1)$ the following degenerate (but ergodic) Markov chain
which also has the equidistribution as its invariant measure.  
{\begin{equation}\label{degenerate}
M= \left(  \begin{matrix} 
     0&1&0\\
     p  &0&1-p\\
      1-p &0&p\\
  \end{matrix}\right)
  \end{equation}}

\begin{figure}[htb]\label{3}
\setlength{\unitlength}{1mm}
\begin{center}
\begin{picture}(125,45)
\put(52.5,33)
{\xymatrix{
        & 1 \ar@/^-0.3cm/[ldd] &   &  \\
         &                    &   &  \\
2 \ar@/^-0.3cm/[rr] \ar@/^-0.3cm/[ruu]   &  & ~3
\ar@/^-0.4cm/[luu] \ar@(ur,dr)[]  & \\
}}
\end{picture}
\end{center}
\caption{The transition graph for matrix \eqref{degenerate}  with a
deterministic transition from state 1 to state 2.}
\end{figure}
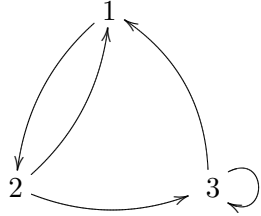

The result for the metastate takes the following surprising form.
\begin{thm} \label{Theorem3} The metastate in the $3$-state random-field Potts model defined above, with a random-field generated  
by the Markov chain $M$ of  eq. \eqref{degenerate}, has the form 
\begin{equation}
\k[\eta]=\frac{1}{2}\d_{\mu^3[\eta]}+\frac{1}{3}\d_{\frac{1}{2}\mu^1[\eta]+ \frac{1}{2}\mu^2[\eta]}
+\frac{1}{9}\d_{p(\b,B)\mu^1[\eta]+ (1-p(\b,B))\mu^2[\eta]}
+\frac{1}{18} \d_{(1-p(\b,B))\mu^1[\eta]+p(\b,B)\mu^2[\eta] }
\end{equation}
Here the function $p(\b,B)$ is computable in terms of the mean-field parameter $u$ 
and is strictly bigger than $1/2$ in the phase transition regime. 
\end{thm}

Note that  the metastate is supported not only on pure states 
but also on mixtures, and moreover the weights are non-symmetric under exchange of the state $1$ and $2$. 

\begin{figure}[htbp]
  \centering
  \includegraphics[height=5cm,width=8cm]{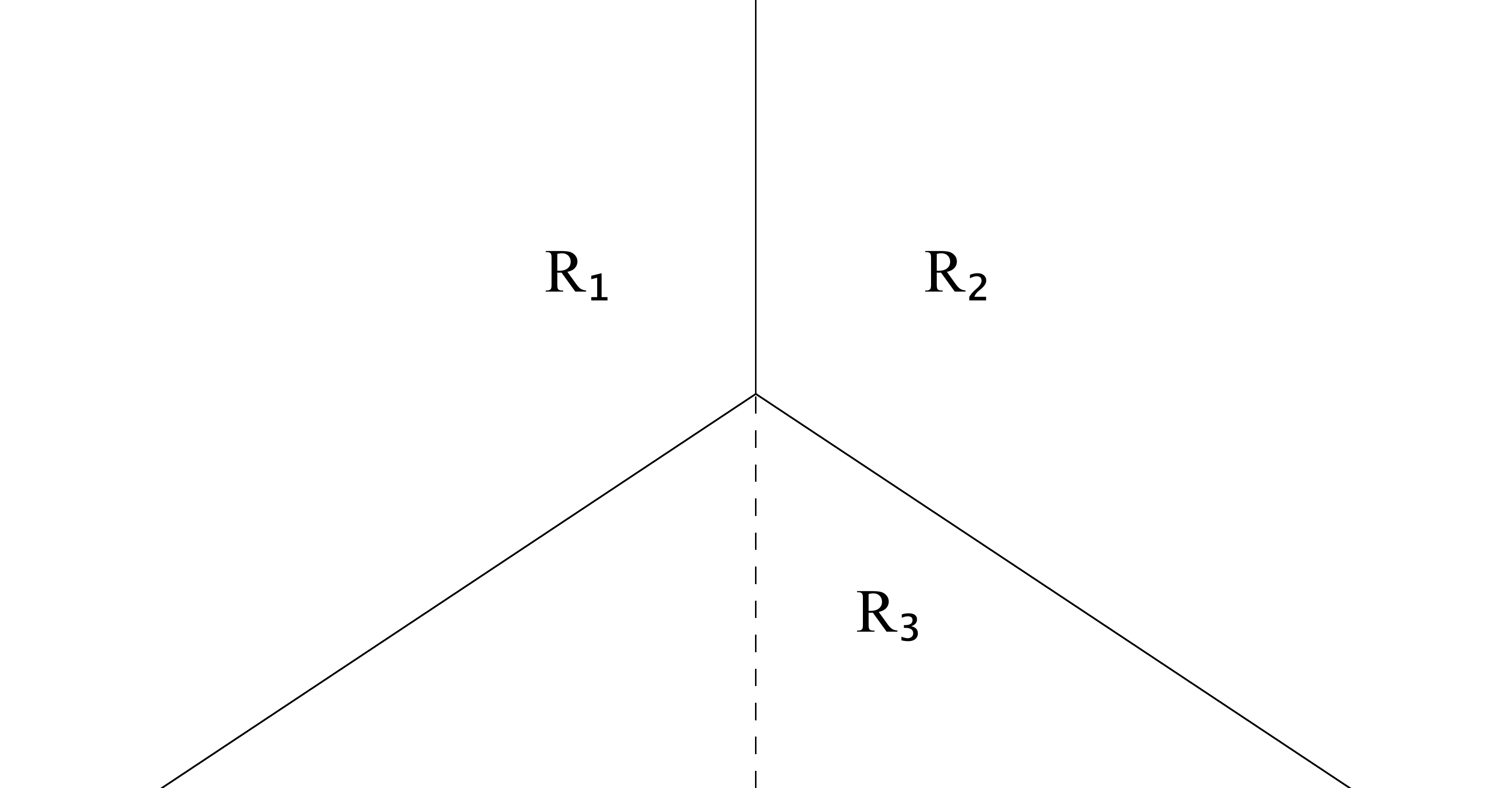} 
  \caption{The Gaussian limiting distribution of $\sqrt{n}(\hat\pi_n-\pi)$ concentrates on the dashed line that for upper half
coincides with the boundary between the stability regions $R_1$ and $R_2$.  
  }
  \label{fig:de}
 \end{figure}

Note that if we tried to apply Theorem \ref{Theorem2} to the case of 
the degenerate chain we would find a Gaussian for the disorder fluctuations in the two-dimensional space 
$T \PP(E')$ which is supported by  the one-dimensional sub-space which contains the 
boundary between the stability regions for the states $\mu^1[\eta]$ and $\mu^2[\eta]$, see  Figure \ref{fig:de}.  
Hence the pure state $\mu^3[\eta]$ should occur with a weight of $1/2$ in the metastate. 
In order to make statements about the occurrence of mixtures of $\mu^1[\eta]$ and $\mu^2[\eta]$ 
and their weights the refined analysis of Section 4 is needed.

\newpage 

\section{Metastates for non-degenerate Markov Chains: Proof }
The proof of Theorem \ref{Theorem1}  in the case of i.i.d. distributed disorder variables relies in an essential way on the product nature of the disorder measure. 
The main step in the present case is a modification of Lemma 2.10 in \cite{CI} to the dependent case. 
For the rest, our argument follows closely the proof of K\"ulske and Iacobelli, and we give details only where the dependence of the disorder variables come into play.
Nevertheless, for the sake of completeness we recall some notation and results from \cite{CI}.\bigskip\\

\noindent Once we have the CLT for the empirical measure, the weights $w_j$ come from the study 
of the limiting probability of the $n,l$-dependent \textit{good sets} $\HH_{i,n,l}^\t$ of the realization of the randomness.
For $n_1<n_2$, let us put
\begin{equation}
X_{[n_1,n_2]}[\eta]=\frac{1}{\sqrt{n_2-n_1+1}}\sum_{i=n_1}^{n_2}\d_{\eta_i}-\sqrt{n_2-n_1+1}\pi
\end{equation}
and define $n,l$-dependent {\em good-sets} $\HH_{i,n,l}^{\t}$ of the realization of the randomness as follows
\begin{equation}
\begin{split}\label{nondeg3}
&\HH_{i,n,l}^{\t}:=\left\lbrace \eta \in (E')^{n-l}: X_{[l+1,n]}[\eta]\in R_{i,n}^{\t}\right\rbrace  \cr 
&\HH_{n,l}^{\t}:=\bigcup_{i=1}^k \HH_{i,n,l}^{\t}
\end{split} 
\end{equation} 
where $R_{i,n}^{\t}:=\{x\in T\PP(E'): \langle x,  B_i \rangle-\max_{k\neq i}    \langle x, B_k \rangle> n^{-\frac{1}{2}+\t}, \Vert x\Vert \leq n^{\frac{\t}{4}} \}$, where \mbox{$0<\t<\frac{1}{2}$}.
The set $\HH_{i,n,l}^{\t}$ of the disorder random variables 
allows to deduce that the measure on the empirical distribution will 
be concentrated inside a ball around the minimizer $\pi \hat\nu^*_i$.  A similar concentration takes place for averages of continuous test-functions with respect to the empirical spin-distribution (see \cite{CI} for a proof).
\begin{lem}\label{concentration}
For any real-valued continuous function $g$ on $\PP(E)$ the following concentration property holds:
\begin{equation}
 |\rho[\eta](n)(g)-g(\pi\hat{\nu}_j^*)|\leq\tilde{r}(n),\quad\forall\,\eta\in\HH_{j,n,0}^{\t},
\end{equation}
where $\lim_{n\rightarrow\infty}\tilde{r}(n)=0$.
\end{lem}
The fundamental property of the good sets is that their probability does not depend on any finite number $l$ of sites in the limit $n\rightarrow\infty$. In order to make this precise, the authors in \cite{CI} considered sligthly different sets. For any $l<n$, we 
define a subregion $\HH_{i,n,0}^{\t}(l)$ of $\HH_{i,n,0}^{\t}$, as follows
\begin{equation}
\begin{split}\label{nondeg5}
&\HH_{i,n,0}^{\t}(l):=
\biggl\{ \eta \in (E')^n:\cr & a_n\langle X_{[1,l]}[\eta],  B_i \rangle+b_n\langle X_{[l+1,n]}[\eta],  B_i \rangle-\max_{k\neq i} \left( a_n\langle X_{[1,l]}[\eta],  B_k \rangle\right) -\max_{k\neq i}\left( b_n\langle X_{[l+1,n]}[\eta],  B_k \rangle\right)>{\delta_n}, \cr & \text{ and } \Vert X_{[1,n]}[\eta]\Vert\leq n^{\frac{\t}{4}}\biggr\} \cr 
\end{split} 
\end{equation} 
where $a_n=\frac{\sqrt{nl}}{n}$ and $b_n=\frac{\sqrt{n(n-l)}}{n}$.\\

\noindent The properties and the relations between $\HH_{i,n,0}^\t$ and $\HH_{i,n,0}^i(l)$ are summarized  in the following lemmas. We refer to \cite{CI} for the details.\\
\noindent Our first
lemma states that for any fixed integer $l$, there is a subregion of the {\em good-set} which will not play any role in the limit $n \uparrow\infty$.
\begin{lem}\label{probforH}
For any integer $l$, $\P(\HH_{i,n,0}^{\t} \setminus \HH_{i,n,0}^{\t}(l))$ goes to zero in the limit $n\uparrow \infty$.
\end{lem}

We also need the sets
\begin{equation}
 \begin{split}\label{partition}
  &B_{i,n}^{{\t}}(l)=(E')^l\times\HH_{i,n,l}^{1,\t}\cr
&A_{i,n}^{{\t}}(l)=(E')^l\times\HH_{i,n,l}^{2,\t}\cr
 \end{split}
\end{equation}

\noindent where
\begin{equation}
\begin{split}\label{region(n-L)}
  &\HH_{i,n,l}^{1,\t}=\biggl\{\eta\in (E')^{n-l}:\phi^i(X_{[l+1,n]}[\eta])>b_n^{-1}\left( {\delta_n} + a_n C_2(l)\right),\text{ and } \Vert X_{[l+1,n]}[\eta]\Vert\leq b_n ^{-1}(n^{\frac{\t}{4}}-\tilde C_2(l))  \bigg\}\cr
&\HH_{i,n,l}^{2,\t}=\biggl\{\eta\in (E')^{n-l}:\phi^i(X_{[l+1,n]}[\eta])>b_n^{-1}\left( {\delta_n} - a_n C_2(l)\right),\text{ and }\Vert X_{[l+1,n]}[\eta]\Vert\leq b_n ^{-1}n^{\frac{\t}{4}}\bigg\}\cr
\end{split}
\end{equation}
with the definitions
\begin{equation}
\phi^i(X_{[l+1,n]}[\eta]):=\langle X_{[l+1,n]}[\eta],B_i\rangle-\max_{k\neq i}\langle X_{[l+1,n]}[\eta],B_k\rangle
\end{equation}
and 
\begin{equation}
\begin{split}
& C_2(l)=\sqrt{l}\max_{\eta}\left|\max_{k\neq i}\langle X_{[1,l]}[\eta],B_k\rangle-\langle X_{[1,l]}[\eta],B_i\rangle\right|
\cr
& \tilde{C}_2(l)=\max_{\eta}||X_{[1,l]}[\eta]||.\cr
\end{split}
\end{equation}

\noindent The following holds:
\begin{enumerate}
\item From the definitions follow
\begin{equation}
 \begin{split}\label{inclusion}
 & A_{i,n}^{{\t}}(l)\supseteq \HH_{i,n,0}^{\t}(l) \supseteq B_{i,n}^{{\t}}(l)\cr
&\HH_{i,n,l}^{1,\t}\subseteq\HH_{i,n,l}^{2,\t}\cr
 \end{split}
\end{equation}
\item For any integer $l$, $\P(\HH_{i,n,l}^{2,\t} \setminus \HH_{i,n,l}^{1,\t})$ goes to zero in the limit $n\uparrow \infty$ \cite{CI}.
\end{enumerate}
The next lemma represents a  fundamental ingredient for what is coming next. It follows from a straightforward modification of Lemma 2.9 in \cite{CI}, when the multidimensional CLT for i.i.d. random variables
is substituted with the one for Markov chains (See Appendix). 

\begin{lem}\label{pesi}
 For any integer $l$, $ \lim_{n\uparrow\infty}\P(\HH_{i,n,l}^{1,\t})=\P_{\Sigma_M}(G\in R_{i})$
where $G\sim\NN(0,\Sigma_M)$.
\end{lem}
\noindent
\vskip .1in
\noindent
At this point we are ready to give the main step for the proof of Theorem \ref{Theorem1} as Lemma \ref{mainstep}. Here the markovian nature of the disorder comes into play.\\ 
First let us now summarize what we have done above for the decompositions of the various regions of the $\eta$-configuration space.
\begin{equation}
\begin{split}\label{nondegh5}
&1_{\HH_{i,n,0}^{\t}}=1_{\HH_{i,n,0}^{\t}(l)}+1_{\HH_{i,n,0}^{\t} \setminus \HH_{i,n,0}^{\t}(l)}\cr
&1_{B_{i,n}^{\t}(l)}=1_{(E')^l}1_{\HH_{i,n,l}^{1,\t}}\cr
&1_{A_{i,n}^{\t}(l)}=1_{(E')^l}1_{\HH_{i,n,l}^{2,\t}}
\end{split}
\end{equation}

\noindent Then, let  be $\psi$ a continuous real-valued function on $\mathcal{P}(\mathcal{P}(E))\times (E')^m$, for some $m>0$,  the key lemma reads.

\begin{lem}\label{mainstep} Under the assumption that the Markov chain disorder has a full rank covariance matrix $\Sigma_M$ and, under non-degeneracy conditions 1. and 2.  on the minimizers  of the free-energy functional (see pag.\pageref{aaa}), it holds
\begin{equation}
\lim_{n\rightarrow\infty}\int_{\HH_{i,n,0}^{\t}}\P_{\pi}(d\eta)\psi(\rho[\eta](n),\eta)=w_i\int_{(E')^m}\P_{\pi}(d\eta(1),\ldots,d\eta(m))\psi(\delta_{\pi\hat{\nu}_i},\eta)
\end{equation}
\end{lem}
\textbf{Proof:} 
We have the decomposition:
\begin{equation}\label{1deco}
\int_{\HH_{i,n,0}^{\t}}\P_{\pi}(d\eta)\psi(\rho[\eta](n),\eta)=\int_{\HH_{i,n,0}^{\t}}\P_{\pi}(d\eta)[\psi(\rho[\eta](n),\eta)-\psi(\delta_{\pi\hat\nu_i},\eta)]+\int_{\HH_{i,n,0}^{\t}}\P_{\pi}(d\eta)\psi(\delta_{\pi\hat\nu_i},\eta).
\end{equation}
We can assume $\psi(\rho,\eta)=\tilde\psi(\rho(g_1),\ldots,\rho(g_l),\eta_{[1,m]})$ for a finite $l$ with continuous $\tilde\psi$ and bounded $g_i$'s. 
Then, because of Lemma \ref{concentration}, the first term on the right-hand side goes to 0 in the limit $n\rightarrow\infty$.\\
Set $l>m$ . The second term can be rewritten as
(in the next we skip the dependence of $\psi$ on $\eta(1),\ldots,\eta(n)$ in the notation)
\begin{equation}\label{2deco}
\int_{\HH_{i,n,0}^{\t}}\P_{\pi}(d\eta)\psi(\delta_{\pi\hat\nu_i})=\int_{\HH_{i,n,0}^{\t}(l)}\P_{\pi}(d\eta)\psi(\delta_{\pi\hat\nu_i})+\int_{\HH_{i,n,0}^{\t}\setminus \HH_{i,n,0}^{\t}(l)}\P_{\pi}(d\eta)\psi(\delta_{\pi\hat\nu_i}),
\end{equation}
where the last term also goes to zero in the limit since $\psi$ is bounded and $\P_{\pi}\left(\HH_{i,n,0}^{\t}\setminus \HH_{i,n,0}^{\t}(l)\right)$ equals zero as $n$ goes to infinity. Moreover, from \eqref{nondegh5}
we have the inequalities:
\begin{equation}\label{ineq}
\int_{B_{i,n}^{\t}(l)}\P_{\pi}(d\eta)\psi(\delta_{\pi\hat\nu_i})\leq\int_{\HH_{i,n,0}^{\t}(l)}\P_{\pi}(d\eta)\psi(\delta_{\pi\hat\nu_i})\leq \int_{A_{i,n}^{\t}(l)}\P_{\pi}(d\eta)\psi(\delta_{\pi\hat\nu_i}).
\end{equation}
Let us consider the last one.  Conditioning on $\eta(1),\ldots,\eta(m)$, and using the Markov property,  we have:
\begin{align*}
&\quad\int_{A_{i,n}^{\t}(l)}\P_{\pi}(d\eta)\psi(\delta_{\pi\hat\nu_i})\\
&=\int_{(E')^m}\left\{\P_{\pi}(d\eta(1),\ldots,d\eta(m))\psi(\delta_{\pi\hat\nu_i})\int_{(E')^{l-m}\times\HH_{i,n,l}^{2,\t}} \P(d\eta(n),\ldots,d\eta(m+1)|d\eta(m))\right\}\\
&=\int_{(E')^m}\big\{\P_{\pi}(d\eta(1),\ldots,d\eta(m))\psi(\delta_{\pi\hat\nu_i})\\
&\quad\times\int_{(E')^{l-m}\times\HH_{i,n,l}^{2,\t}} \P(d\eta(n),\ldots,d\eta(l+1)|d\eta(l))\P(d\eta(l),\ldots,d\eta(m+1)|d\eta(m))\big\}\\
&=\int_{(E')^m}\big\{\P_{\pi}(d\eta(1),\ldots,d\eta(m))\psi(\delta_{\pi\hat\nu_i})\\
&\quad\times\int_{(E')\times\HH_{i,n,l}^{2,\t}} \P(d\eta(n),\ldots,d\eta(l+1)|d\eta(l))\int_{(E')^{l-1-m}}\P(d\eta(l),\ldots,d\eta(m+1)|d\eta(m))\big\}\\
&=\int_{(E')^m}\big\{\P_{\pi}(d\eta(1),\ldots,d\eta(m))\psi(\delta_{\pi\hat\nu_i})\\
&\quad\times\int_{(E')\times\HH_{i,n,l}^{2,\t}} \P(d\eta(n),\ldots,d\eta(l+1)|d\eta(l))\P(d\eta(l)|d\eta(m))\big\}.\\
\intertext{From the Perron-Frobenius theorem (and its consequences), as presented for example in \cite{Bre98}, it follows $\P(d\eta(l)|d\eta(m))=\pi(\eta(l))+O((l-m)^{m_2-1}\mu^{l-m})$ where $\mu$ is the second largest eigenvalue of $M$ in modulus and $m_2$ is the algebraic multiplicity of $\mu$. Thus the last integral becomes:}
&=\int_{(E')^m}\big\{\P_{\pi}(d\eta(1),\ldots,d\eta(m))\psi(\delta_{\pi\hat\nu_i})\\
&\quad\times\int_{(E')\times\HH_{i,n,l}^{2,\t}} \P(d\eta(n),\ldots,d\eta(l+1)|d\eta(l))\pi(\eta(l))\big\}+O((l-m)^{m_2-1}\mu^{l-m})\\
&=\int_{(E')^m}\P_{\pi}(d\eta(1),\ldots,d\eta(m))\psi(\delta_{\pi\hat\nu_i})\int_{\HH_{i,n,l}^{2,\t}} \P_{\pi}(d\eta(n),\ldots,d\eta(l+1))+O((l-m)^{m_2-1}\mu^{l-m}).
\end{align*}
A similar computation holds for the left-hand side of \eqref{ineq}. Recall that $|\mu|<1$. Taking for both the limit $n\rightarrow \infty$ first and then $l\rightarrow\infty$
we obtain the result.
$\Cox$\bigskip\\
We are now ready to complete the proof of the theorem. For any bounded function $\psi$ we write,

\begin{align}\label{222}
\int \P_{\pi}(d\eta)\psi(\rho[\eta](n),\eta)=&\sum_{i=1}^k \int \P_{\pi}(d\eta)\psi(\rho[\eta](n),\eta)1_{\mathcal{H}^{\tau}_{i,n,0}}(\eta)\\
&+ \int \P_{\pi}(d\eta)\psi(\rho[\eta](n),\eta)1_{(\mathcal{H}^{\tau}_{i,n,0})^c}(\eta)
\end{align}
Here the sum runs over the minimizers of the free energy.
The second non-degeneracy assumption ensures that the second term of formula \eqref{222} goes to 0 in the limit as $n\rightarrow\infty$ \cite{CI}. 
Then applying Lemma \ref{mainstep} 
we have
 \begin{equation}\label{225}
\lim_{n\rightarrow\infty}\int \P_{\pi}(d\eta)\psi(\rho[\eta](n),\eta)=\sum_{i=1}^k \int \P_{\pi}(d\eta)\psi(\rho[\eta](n),\eta)w_i\delta_{\delta_{\pi\hat\nu_i}}(d\rho)
\end{equation}
which by the definition of AW-metastate gives us
\begin{equation}\label{223}
K(d\rho,d\eta)=\sum_{i=1}^k\mathbb{P}_{\pi}(d\eta)w_i\delta_{\delta_{\pi\hat\nu_i}}(d\rho)
\end{equation}
and 
\begin{equation}\label{224}
K[\eta](d\rho)=\sum_{i=1}^k\omega_i\delta_{\delta_{\pi\hat\nu_i}}(d\rho).
\end{equation}
This concludes the proof of Theorem \ref{Theorem1}.
$\Cox$
\bigskip 
\bigskip

After we have been quite precise in the previous treatment, let us for 
the proof of  Theorem \ref{Theorem2}  go a little faster.  {Since any continuous function on $\mathcal{P}(E^{\infty})\times (E')^{\infty}$ has a finite-dimensional approximation, it suffices to consider
a local function $F$ which depends on $m$ coordinates of spins and random fields. For such an $F$ we need to prove that}
\begin{equation*}
\lim_{n\uparrow\infty}\int_{\HH_{j,n}^{\delta_n}}\P_\pi(d\eta)F(\mu_n[\eta],\eta)=w_j\int_{(E')^m} 
{\mathbb{P}_{\pi}}
(d\eta)  
F\Bigl( \prod_{i=1}^m \g[\eta(i)](\,\cdot\,| \pi \hat\nu_j),\eta\Bigr)  
\end{equation*}
This is done using the asymptotic decoupling of the occupation times from 
any starting configuration, allowing for a little extra margin in volume to account 
for exponential memory loss along the path of the Markov chain. {The weights $w_j$ come out from the analysis performed in Lemma \ref{mainstep} 
exploiting the (exponentially) fast convergence to equilibrium of the ergodic Markov chain providing the disorder field. Together with this we use the concentration (for $n$ large and $\eta\in \HH_{j,n}^{\t}$)
of $\mu_{F,n}[\eta](A)$ around $\mu_j[\eta](A)$ where  $\mu_j[\eta]:=\prod_{i=1}^m \g[\eta(i)](\,\cdot\,| \pi \hat\nu_j)$ and $A$ is an event depending on the first $m$ coordinates \cite{CI}.
With this in hand,}
the proof is parallel to the independent case. 
$\Cox$

\newpage 

\section{Random field Potts model, degenerate Markov chain}

The entries of the covariance matrix  
$\S_{M,n}$ of the standardized occupation time vector in finite volume $n$ are  
\begin{allowdisplaybreaks}
\begin{align*}
&\S_{M,n}(i,j)=n\,\,\E_{\pi}\Bigl((\hat\pi_n(i)-\pi(i))(\hat\pi_n(j)-\pi(j))\Bigr)\\
&=\pi(i)\cdot1_{\{i=j\}}-\pi(i)\pi(j)+\frac{\pi(i)}{n}\sum_{r=1}^{n-1}(n-r)\left[M^r(i,j)-\pi(j)\right]\\
&+\frac{\pi(j)}{n}\sum_{r=1}^{n-1}(n-r)\left[M^r(j,i)-\pi(i)\right].
\end{align*}
\end{allowdisplaybreaks}

Consider the case $q=3$ of a general doubly stochastic matrix 
in the form  
\begin{equation}
  M=\left( \begin{matrix} 
      a & b&1-a-b \\
      c & d&1-c-d \\
      1-a-c & 1-b-d&-1+a+b+c+d \\
   \end{matrix}\right),\quad a,b,c,d\in(0,1).
\end{equation}
The limit when $n$ tends to infinity of the covariance becomes 
\begin{equation}\label{Pottsweights}
\S_M= \left(  \begin{matrix} 
     \frac{2}{9} + \frac{2(1+b(2-6c)+2c-2d+a(-5+6d))}{27(-1+a+bc+d-ad)} & -\frac{1}{9} - \frac{b(5-6c)+5c-2(1+d)+a(-2+6d)}{27(-1+a+bc+d-ad)} & -\frac{1}{9} - \frac{4-8a-b-c-6bc-2d+6ad}{27(-1+a+bc+d-ad)}\\
     &&\\
        -\frac{1}{9} - \frac{b(5-6c)+5c-2(1+d)+a(-2+6d)}{27(-1+a+b+c+d-ad)} &  \frac{2}{9} + \frac{2(1+b(2-6c)+2c-5d+a(-2+6d))}{27(-1+a+b+c+d-ad)} &-\frac{1}{9} - \frac{4-2a-b-c-6bc-8d+6ad}{27(-1+a+bc+d-ad)} \\
    &&    \\
        -\frac{1}{9} - \frac{4-8a-b-c-6bc-2d+6ad}{27(-1+a+bc+d-ad)}&     -\frac{1}{9} - \frac{4-2a-b-c-6bc-8d+6ad}{27(-1+a+bc+d-ad)} & \frac{2}{9} - \frac{2(-4+b+c+6bc+a(5-6d)+5d)}{27(-1+a+bc+d-ad)}  \\
   \end{matrix}\right)
\end{equation}
which still depends on $a,b,c$ and $d$.\bigskip\\
Indeed this can be done invoking the  geometric series and diagonalization which is a $q-1=2$ dimensional problem since 
we always have the eigenvalue $1$, using some help of Mathematica. 

The restriction of $\S_M$ to the two dimensional space $T \PP(E')$ gives 
us the weights $w_j$ in the Potts-metastate (\ref{Pottsmeta}) which are generically different. 
We note that the degenerate transition matrix from the last part of Chapter 2 also 
falls into this class. Let us now provide the proof of the last theorem of the introduction. 
\bigskip 

{\bf Proof of Theorem \ref{Theorem3}. } Consider at first the chain 
which starts in $\eta(1)=3$, and use the corresponding definition of a metastates with this starting measure, 
calling it $\k_3[\eta]$.  
We then have for the difference of occupation times 
$n \hat \pi_n(1) - n \hat  \pi_n(2)=l=0,1$ as the only two possible values along a path $\eta$ with length $n$ 
with non-zero probability. 
More precisely we have $l=1$ iff $\eta(n)=1$ and $l=0$ iff $\eta(n)\in \{2,3\}$. Call this property $(\star)$.


Let us condition ourselves at first on the event of  MC paths where $\eta(n)=1$  and $\hat \pi_n(3)$ is NOT $3$-like. 
(By the latter we mean that it does not lie in the stability region for the state $3$, with an $n$-dependent safety zone 
around it, as defined in Section 3.)  
This means that conditional on this disorder event 
we do not see the 3-state, we see asymptotically a mixture $p(\b,B)\mu^1[\eta]+ (1-p(\b,B))\mu^2[\eta]$ 
between the 1-state and the 2-state 
with a bias $p(\b,B)$ for the 1-state which we expect to be bigger than $1/2$.  
To determine this weight note that $p(\b,B)/(1-p(\b,B))$ will be asymptotically given by 
\begin{equation}\label{4.343}
\begin{split}
&\frac{\displaystyle\sum_{\s: \Vert L_n(\s)- \pi\hat\nu^*_1 \Vert \leq \e} \mu_{F,n}[\eta(1),\dots,\eta(n-1),\eta(n)=1]( \s)}
{\displaystyle\sum_{\s: \Vert L_n(\s)- \pi\hat\nu^*_2 \Vert \leq \e} \mu_{F,n}[\eta(1),\dots,\eta(n-1),\eta(n)=1]( \s)}\cr
\end{split}
\end{equation}
To compute the limit of this expression we use finite-volume manipulations to obtain 
a conditional Gibbs expectation in the following way.  
We use the property $(\star)$ and rewrite \eqref{4.343}, for a possible path $ \eta(1),\dots,\eta(n-1),\eta(n)$ of the chain, 
\begin{equation}\label{4.3}
\begin{split}
&\frac{\displaystyle\sum_{\s: \Vert L_n(\s)- \pi\hat\nu^*_2 \Vert \leq \e} \mu_{F,n}[\eta(1),\dots,\eta(n-1),\eta(n)=2](\s)}
{\displaystyle\sum_{\s: \Vert L_n(\s)- \pi\hat\nu^*_2 \Vert \leq \e} \mu_{F,n}[\eta(1),\dots,\eta(n-1),\eta(n)=1]( \s)}\cr
\end{split}
\end{equation}
We warn the reader that the $\eta$-sequence appearing in the Gibbs measure 
in the numerator is not an allowed sequence of the Markov chain any more, but nevertheless 
the Gibbs expression is well defined. 

Splitting off the Hamiltonian appearing in the denominator gives us a difference term in the exponent of the numerator 
at the site $n$, and we can rewrite the quotient \eqref{4.3} as a fraction of partition functions 
with one random field value changed and a conditional Gibbs measure in the form 
\begin{equation}
\begin{split}\label{fourfive}
&\frac{Z_{F,n}[\eta(1),\dots,\eta(n-1),\eta(n)=1]}{Z_{F,n}[\eta(1),\dots,\eta(n-1),\eta(n)=2]}\cr
&\times \Bigl(\mu_{F,n}[\eta(1),\dots,\eta(n-1),\eta(n)=2](  e^{B (1_{\s_n=1}- 1_{\s_n=2})}|\Vert L_n(\s)- \pi\hat\nu^*_2 \Vert \leq \e )\Bigr)^{-1}\cr
\cr
\end{split}
\end{equation}
Note that, by symmetry we have the equality  
\begin{equation}
\begin{split}
&\frac{Z_{F,n}[\eta(1),\dots,\eta(n-1),\eta(n)=1]}{Z_{F,n}[\eta(1),\dots,\eta(n-1),\eta(n)=2]}=1\cr
\end{split}
\end{equation}
Using
\begin{equation}
\begin{split}
& e^{B (1_{\s_n=1}- 1_{\s_n=2})}
=(e^{B}-1)1_{\s_n=1}+(e^{-B}-1)1_{\s_n=2}+1
\end{split}
\end{equation}
the expected value in \eqref{fourfive} can be computed, where we note 
the constraint $\eta(n)=2$, and in this way we get 
\begin{equation}
\begin{split}
&\mu_{F,n}[\eta(1),\dots,\eta(n-1),\eta(n)=2](  e^{B (1_{\s_n=1}- 1_{\s_n=2})}|\Vert L_n(\s)- \pi\hat\nu^*_2 \Vert \leq \e )\cr
&=(e^{B}-1)
\mu_{F,n}[\eta(1),\dots,\eta(n-1),\eta(n)=2]( 1_{\s_n=1} |\Vert L_n(\s)- \pi\hat\nu^*_2 \Vert \leq \e )\cr
&\qquad+(e^{-B}-1)\mu_{F,n}[\eta(1),\dots,\eta(n-1),\eta(n)=2]( 1_{\s_n=2} |\Vert L_n(\s)- \pi\hat\nu^*_2 \Vert \leq \e )+1
\end{split}
\end{equation}
Taking into account the conditioning of the total empirical vector and the local knowledge of the random field we have e.g. 
\begin{equation}
\begin{split}
&\mu_{F,n}[\eta(1),\dots,\eta(n-1),\eta(n)=2]( 1_{\s_n=1} |\Vert L_n(\s)- \pi\hat\nu^*_2 \Vert \leq \e )\cr
&\rightarrow \g[\eta_n=2](1_{\s_n=1} | \pi\hat\nu^*_2) 
\end{split}
\end{equation}
This gives 

\begin{equation}
\begin{split}
&\mu_{F,n}[\eta(1),\dots,\eta(n-1),\eta(n)=2](  e^{B (1_{\s_n=1}- 1_{\s_n=2})}|\Vert L_n(\s)- \pi\hat\nu^*_2 \Vert \leq \e )\cr
&\rightarrow (e^{\b \e}-1) \g[\eta_n=2](1_{\s_n=1} | \pi\hat\nu^*_2) 
+(e^{-B}-1)\g[\eta_n=2](1_{\s_n=2} | \pi\hat\nu^*_2) +1 =: p^{-1}_1(\b,B)
\end{split}
\end{equation}
With the relation $\frac{p(\b,B)}{1-p(\b,B)}=p_1(\b,B)$  the metastate for the chain started in state $3$ becomes
\begin{equation}
\begin{split}
&\k_3[\eta]=\lim_n \Biggl( \P(\hat \pi_n \text{ is 3-like})\d_{\mu^3[\eta]}
+  \P(\hat \pi_n \text{ is not 3-like and } \eta_n  \in \{2,3\}) \d_{\frac{1}{2}\mu^1[\eta]+ \frac{1}{2}\mu^2[\eta]}\cr
&+  \P(\hat \pi_n \text{ is not 3-like and } \eta_n  =1) \d_{p(\b,\e)\mu^1[\eta]+ (1-p(\b,B))\mu^2[\eta]}\Biggr)\cr
&=\frac{1}{2}\d_{\mu^3[\eta]}
+\frac{1}{3}\d_{\frac{1}{2}\mu^1[\eta]+ \frac{1}{2}\mu^2[\eta]}+\frac{1}{6} \d_{p(\b,B)\mu^1[\eta]+ (1-p(\b,B))\mu^2[\eta]}\cr
\end{split}
\end{equation}
In the second line we have used that for $\eta_n  \in \{2,3\}$ the states $1$ and $2$ are equivalent. 
Further we have used the CLT of the form of Theorem \ref{tha} (see the Appendix) and the asymptotic decoupling of the state and the occupation time measure. 
\bigskip\\
The outcome of the previous computation for the metastate depends on the starting point of the Markov Chain. We started with $\eta(1)=3$ and we called $\k_3[\eta]$ the associated metastate. Similar arguments 
give for the metastates where the Markov chain starts in the states $1$ and $2$  imply tha
that $\k_1[\eta]=\k_3[\eta]$ whereas 
\begin{equation}
\k_2[\eta]=\frac{1}{2}\d_{\mu^3[\eta]}
+\frac{1}{3}\d_{\frac{1}{2}\mu^1[\eta]+ \frac{1}{2}\mu^2[\eta]}+\frac{1}{6} \d_{(1-p(\b,B))\mu^1[\eta]+p(\b,B)\mu^2[\eta] }
\end{equation}
When the disorder starts in equilibrium then $\pi(i)=1/3$, $i=1,2,3$, the complete expression for the metastate reads:
\begin{equation}
\k[\eta]=\sum_{i=1}^3\pi(i)\k_i[\eta]=\frac{2}{3}\k_1[\eta]+\frac{1}{3}\k_2[\eta]
\end{equation}
and from this the statement of Theorem \ref{Theorem3} follows.  

Finally let us give some information about the weights $p(\b,B)$. Using the parametrization of the minimizers in terms of the order parameter 
$u$ and the form of the $\g$-kernels of the Potts model we have 
\begin{equation}\label{52}
p^{-1}_1(\b,B)
=\frac{e^B+e^{\beta u}+1}{2+e^{\beta u+B}}
\end{equation}
\noindent Notice that $p$ cannot be equal to $1/2$ in the phase trasition region. 
We claim \eqref{52} is always strictly less than 1 unless $u=0$.  To see this put $x=e^B$ and $y=e^{\beta u}$, then we have to prove that
\begin{equation}
x+y\leq 1+xy\mbox{ with } x>1,y\geq 1.
\end{equation}
This is clear since, for every $x$, we have equality in $y=1$ and the derivative in $y$ of the left term is strictly less than the same derivative on the right.  The equality holds only if $y=1$ or equivalently only if $u=0$.
$\Cox$

\section{Appendix. }
One way to derive a  CLT for $\hat\pi_n$ is to use 
 the results in \cite{KEWI} for one-dimensional processes defined on a finite state Markov chain together with the Cram\'{e}r-Wold device.\\

\begin{thm}\label{tha} Under our hypothesis on the chain $\left(\eta(n)\right)_{n\in\mathbb{N}}$, the following CLT holds \cite{KEWI}:
$$\left(\eta(n), \sqrt{n}(\langle\lambda,\hat\pi_n-\pi\rangle\right)\xrightarrow{\mathcal{L}}\pi\otimes\mathcal{N}(0,\langle\lambda,\Sigma_M\lambda\rangle).$$ 
where $\Sigma_M=\lim_{n\rightarrow +\infty}\Sigma_{M,n}$, with $\Sigma_{M,n}(i,j)= n\times \mbox{Cov}_{\pi}( \hat\pi_n(i),\hat\pi_n(j))$.
\end{thm}
The theorem can also be rephrased saying that the Markov chain $\left(\eta(n)\right)_{n\in\mathbb{N}}$ and the process  $(n\langle\lambda,\hat\pi_n\rangle)_{n\in\mathbb{N}}$ are asymptotically independent.
The theorem is a special case of  Theorem 1.1 of \cite{KEWI} when we
put $Y_n:=n\langle\lambda,\hat\pi_n\rangle$ and $dM(i,j)(y):=M(i,j)\d_{\l(j)}(dy)$ in their paper.   
\\
To prove the multidimensional CLT 
use the previous theorem together with the Cram\'{e}r-Wold device.  
The latter states that a sequence of random vectors $X_1,\ldots, X_n$ converges in distribution to $X$ if and only if 
the scalar product $<\lambda,X_n>$ converges in distribution to  $<\lambda,X>$ for all $\lambda$, $|\lambda|\neq 0$.\\

As an(other) easy way to prove such a result use 
the regeneration structure \cite{Bre98} of the Markov chain, and consider the occupation times 
between recurrences to a fixed reference state. These form an i.i.d. sequence plus a remainder 
which can be controlled, hence the CLT follows.  

\subsection*{Acknowledgements}
We acknowledge support by the Sonderforschungsbereich 
SFB | TR12 - Symmetries and Universality in Mesoscopic Systems 
and the University of  Bochum.\\
The authors also thank Aernout van Enter 
and Giulio Iacobelli 
for interesting discussions and a critical reading of the
manuscript.

\end{document}